\documentclass[10pt,conference]{IEEEtran}
\usepackage{times}
\usepackage[cmex10]{amsmath}
\usepackage{amssymb}
\usepackage{epsfig,verbatim}
\usepackage{algorithm}
\usepackage{algpseudocode}

\newcommand\blfootnote[1]{%
  \begingroup
  \renewcommand\thefootnote{}\footnote{#1}%
  \addtocounter{footnote}{-1}%
  \endgroup
}

\long\def\comment#1{}

\newfont{\bbb}{msbm10 scaled 700}

\newfont{\bb}{msbm10 scaled 1100}

\newcommand{\RR}{\mbox{\bb R}}

\newcommand{\av}{{\bf a}}
\newcommand{\bv}{{\bf b}}
\newcommand{\cv}{{\bf c}}

\newcommand{\hv}{{\bf h}}

\newcommand{\mv}{{\bf m}}

\newcommand{\rv}{{\bf r}}

\newcommand{\tv}{{\bf t}}

\newcommand{\xv}{{\bf x}}
\newcommand{\yv}{{\bf y}}
\newcommand{\zv}{{\bf z}}

\newcommand{\Hm}{{\bf H}}

\newcommand{\Ac}{{\cal A}}

\newcommand{\Cc}{{\cal C}}
\newcommand{\Dc}{{\cal D}}
\newcommand{\Ec}{{\cal E}}

\newcommand{\Hc}{{\cal H}}

\newcommand{\Mc}{{\cal M}}
\newcommand{\Nc}{{\cal N}}

\newcommand{\transp}{{\sf T}}

\newtheorem{definition}{Definition}

\newtheorem{remark}{Remark}

\newcommand{\argmin}{\operatornamewithlimits{argmin}}

\title{Uplink Multiuser Massive MIMO Systems with One-Bit ADCs: A Coding-Theoretic Viewpoint}

\author{
\IEEEauthorblockN{
              Seonho Kim\authorrefmark{1},  Namyoon Lee\authorrefmark{2} and Songnam Hong\authorrefmark{1}}
\IEEEauthorblockA{\authorrefmark{1}Ajou University, Suwon, Korea,\\
              email: \{shkim1005, snhong\}@ajou.ac.kr}
\IEEEauthorblockA{\authorrefmark{2}POSTECH, Pohang, Korea,\\
              email: nylee@postech.ac.kr}
}

\begin{document}

\maketitle

\date{}

\blfootnote{
}

\begin{abstract}
This paper investigates an uplink multiuser massive multiple-input multiple-output (MIMO) system with one-bit analog-to-digital converters (ADCs), in which $K$ users with a single-antenna communicate with one base station (BS) with $n_r$ antennas. In this system, we propose a novel MIMO detection framework, which is inspired by coding theory. The key idea of the proposed framework is to create a non-linear code $\Cc$ of length $n_r$ and rate $K/n_r$ using the encoding function that is completely characterized by a non-linear MIMO channel matrix. From this, a multiuser MIMO detection problem is converted into an equivalent channel coding problem, in which a codeword of the $\Cc$ is sent over $n_r$ parallel binary symmetric channels, each with different crossover probabilities. Levereging this framework, we develop a maximum likelihood decoding method, and show that the minimum distance of the $\Cc$ is strongly related to a diversity order. Furthermore, we propose a practical implementation method of the proposed framework when the channel state information is not known to the BS. The proposed method is to estimate the code $\Cc$ at the BS using a training sequence. Then, the proposed {\em weighted} minimum distance decoding is applied. Simulations results show that the proposed method almost achieves an ideal performance with a reasonable training overhead.
\end{abstract}

\section{Introduction}

Massive multiple-input multiple-output (MIMO) is a promising multiuser MIMO technology for  beyond 5G cellular systems \cite{Larsson}. In massive MIMO systems, the number of antennas at the base station (BS) is scaled up by several orders of magnitude compared to traditional MIMO systems to increase the capacity and energy efficiency \cite{Larsson, Marzetta}. The use of a large number of antennas at the BS, however, considerably increases the hardware cost and the radio-frequency (RF) circuit power consumption \cite{Yang}. To resolve this problem, the use of low-resolution analog-to-digital converters (ADCs) for massive MIMO systems has received increasing attention over the past years \cite{Mezghani,Mo,Mo2}. The one-bit ADC is particularly attractive because there is no need for an automatic gain controller; this reduces the hardware complexity significantly. 

Recently, there have been several works on detection algorithms for massive MIMO systems with low-resolution ADCs. In \cite{Risi,Jacobsson}, simple linear detection methods with one-bit ADCs were shown to provide reasonable symbol-error performance when QPSK and 16-QAM input constellations are used, respectively. An uplink multiuser massive MIMO detector was developed when multi-bit ADCs \cite{Wang} and the spatial modulation transmission technique \cite{Renzo} are used. In \cite{Wang2} and \cite{Wang3},  a multiuser detector using a message-passing algorithm was proposed for a general input symbol in uplink massive MIMO systems with one-bit ADCs. In \cite{Wang4} and  \cite{Choi}, a multiuser detector using low-resolution ADCs was proposed by using convex optimization techniques.


In this paper, we present a novel multiuser MIMO detection framework inspired by coding theory. The proposed framework is to convert a multiuser MIMO detection problem into a channel coding problem, by viewing the concatenation of a channel transformation and one-bit ADC quantization as an encoding function. Specifically, we create a non-linear code $\Cc$ with length $n_r$ and rate $K/n_r$ using the encoding function that is determined by a non-linear MIMO channel matrix. This allows us to see the equivalent coding problem, in which a codeword of the code $\Cc$ is transmitted over $n_r$ parallel binary symmetric channels (BSCs), each with different crossover probabilities. Assuming that perfect channel state information at receiver (CSIR) is available, we consider two types of decoding: 1) minimum distance (MD) and 2) maximum likelihood (ML) decoding. Although MD decoding is optimal in a classical coding problem, it is shown that ML decoding significantly outperforms MD decoding in our problem due to the different channel reliabilities. Using the proposed ML decoding method, it is shown that the minimum distance of the $\Cc$ determines the slope of bit-error probability (BER) (i.e., the diversity order). Furthermore, we present a practical decoding method when CSIR is not available to the BS. By sending a training sequence, the BS estimates a code $\Cc$ without knowing CSIR. With this estimated code, we propose a {\em weighted} MD decoding method, in which the weights are computed using an estimated channel reliability from a training sequence. Simulations results show that the proposed method almost achieves an ideal performance with a reasonable training overhead.

\textbf{Notation:} Lower and upper boldface letters represent column vectors and matrices, respectively. For any vector $\xv$, $d_w(\xv)$ denotes the Hamming weight, i.e., the number of non-zero values in $\xv$. For any two vector $\xv$ and $\yv$, $d_h(\xv,\yv)$ represents the Hamming distance, i.e., the number of positions at which the corresponding symbols are different. To simplify the notation, let $[K] = \{1,\ldots,K\}$ for any positive integer  $K$. 
For any $k \in \{0,1,\ldots,K-1\}$, we let $g(k)=[b_0,b_1,\ldots,b_{K-1}]^{\transp}$ denote the binary expansion of $k$ where 
$k=b_02^0+\cdots+b_{K-1}2^{K-1}$.
 We also let $g^{-1}(\cdot)$ denote its inverse function.

\begin{figure}
\centerline{\includegraphics[width=9cm]{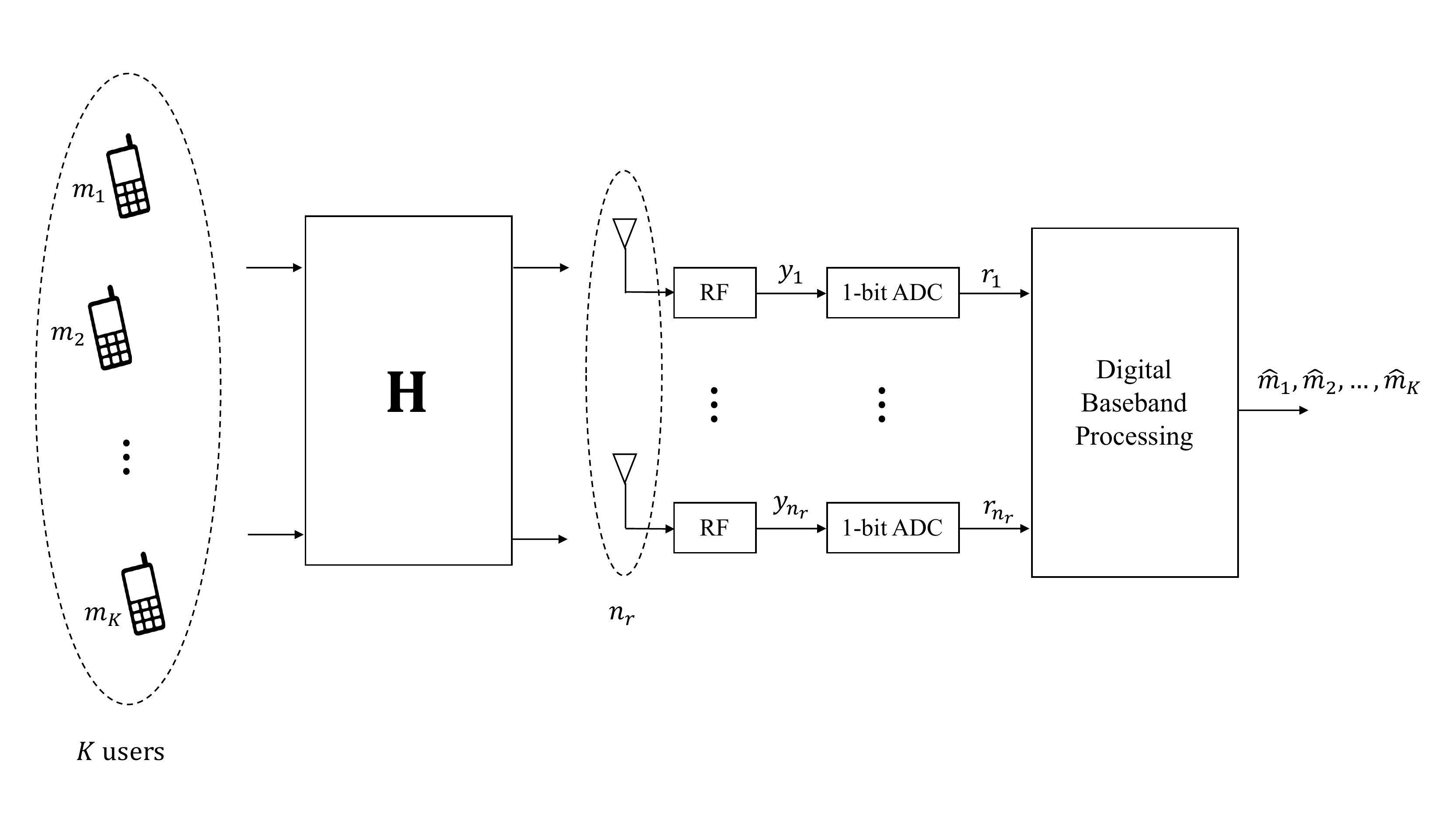}}
\caption{Illustration of an uplink multiuser massive MIMO system in which each receiver antenna at a BS is equipped with one-bit ADC.}
\label{model}
\end{figure}

\section{System Model}\label{sec:model}\label{sec:SM}

We consider a single-cell uplink where there are $K$ single-antenna users and one base station (BS) with an array of $n_r > K$ antennas. The discrete-time real-valued baseband received signal at the BS is 
\begin{equation}
\yv = \Hm\xv +\zv,
\end{equation} where $\Hm \in \RR^{n_r \times K}$ is the channel matrix between the BS and the $K$ users, i.e., $\hv_{i}^{\transp} \in \RR^{1 \times K}$ is the channel vector between the $i$-th receiver antenna at the BS and the $K$ users. The vector $\xv=[x_1,\ldots,x_K]^{\transp}$ contains the transmitted symbols from all the $K$ users. In particular, the $k$-th element of $\xv$, $x_k$, represents the symbol sent by the user $k$. The transmit power constraint is given by $|x_k|^2 \leq \mbox{SNR}$ for $k \in [K]$, and it is assumed that $x_k$ belongs to a BPSK constellation, i.e., $x_k \in \{\sqrt{\mbox{SNR}},-\sqrt{\mbox{SNR}}\}$. Let $m_k \in \{0,1\}$ denote the user $k$'s message. The channel input $x_k$ is obtained by a modulation function $\Mc$ as
\begin{equation}
x_k = \Mc(m_k),
\end{equation} where $\Mc(0) = \sqrt{\mbox{SNR}}$ and $\Mc(1) = - \sqrt{\mbox{SNR}}$. For a vector, the modulation function $\Mc(\cdot)$ is applied element-wise. The elements of noise vector $\zv=[z_1,\ldots,z_{n_r}]^{\transp}$ are independently identically distributed (IID)  Gaussian random variables with zero-mean and unit-variance, i.e., $z_i \sim \Nc(0,1)$.

Let $\mbox{sign}(\cdot): \RR \rightarrow \{0,1\}$ represent the one-bit ADC quantizer function with
\begin{equation}
\mbox{sign}(u)=
\begin{cases}
0 & \mbox{ if } u \geq 0\\
1 & \mbox{ if } u  < 0.
\end{cases}
\end{equation} For a vector, it is applied element-wise. After applying ADC quantizers, the BS observes the quantized received signal as
\begin{equation}
\rv = \mbox{sign}(\yv) \in \{0,1\}^{n_r}.\label{eq:obs}
\end{equation}

In this paper, we only consider a real-valued channel for the ease of understanding of the proposed coding method but it can be straightforwardly applied to a complex-valued channel using the real-valued representation for complex vectors as
\begin{equation}
\left[ {\begin{array}{c}
   \mbox{Re}(\yv)  \\      
   \mbox{Im}(\yv) \\
 \end{array} } \right]=\left[ {\begin{array}{cc}
   \mbox{Re}(\Hm) & -\mbox{Im}(\Hm) \\      
   \mbox{Im}(\Hm) & \mbox{Re}(\Hm)\\
 \end{array} } \right]\left[ {\begin{array}{c}
   \mbox{Re}(\xv)  \\      
   \mbox{Im}(\xv) \\
 \end{array} } \right]+\left[ {\begin{array}{c}
   \mbox{Re}(\zv)  \\      
   \mbox{Im}(\zv) \\
 \end{array} } \right],
\end{equation}where $\mbox{Re}(\av)$ and $\mbox{Im}(\av)$ denote the real and complex part of a complex vector $\av$, respectively.

\begin{figure}
\centerline{\includegraphics[width=8cm]{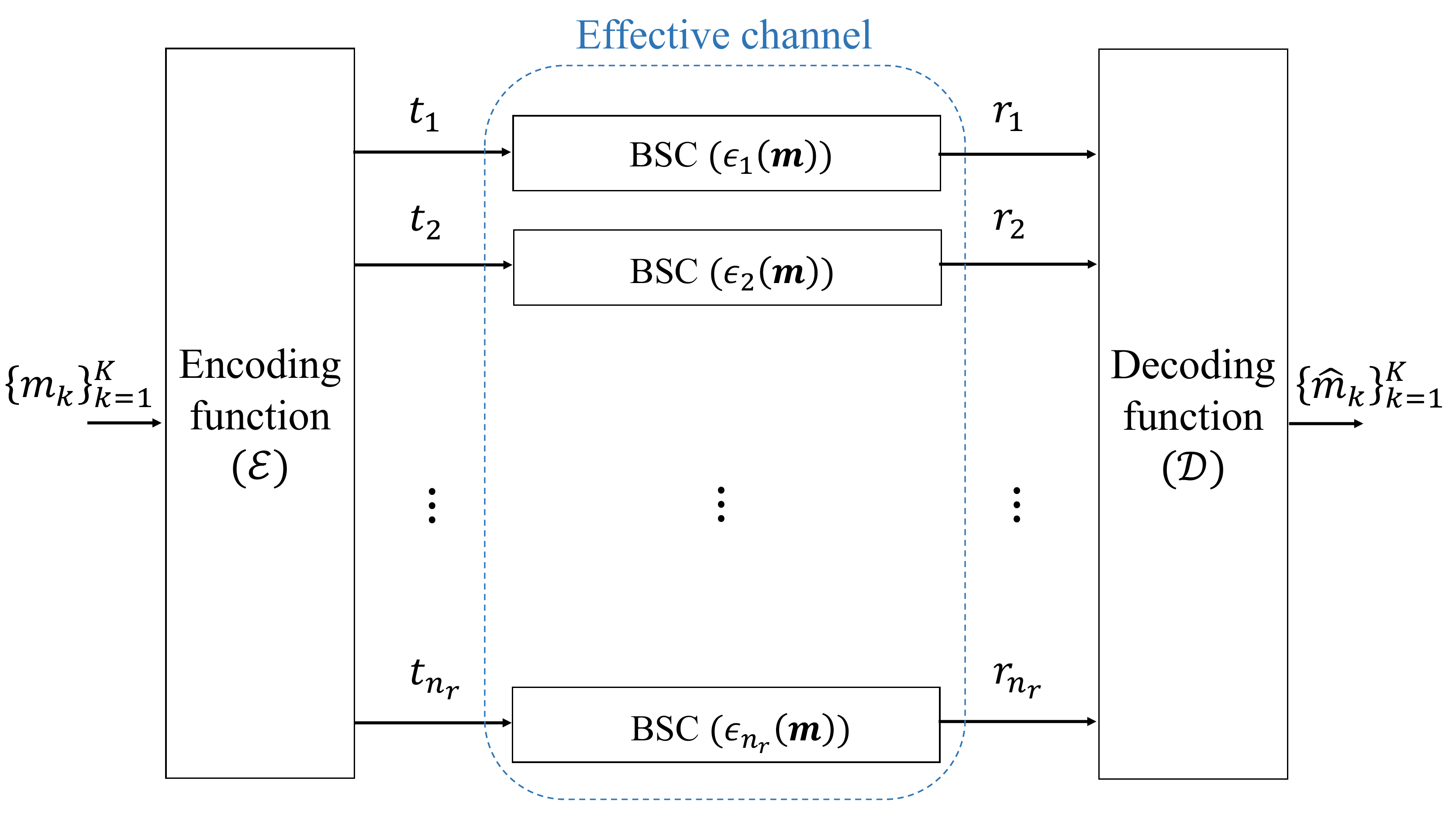}}
\caption{Illustration of an effective communication model to be used for the proposed coding method. Notice that an encoding function $\Ec$ is fixed as a function of $\Hm$ and a decoding function $\Dc$ is proposed.}
\label{e_model}
\end{figure}

\section{The Proposed MIMO Detection Method\\ Using Coding Theory}\label{sec:Main}

In this section, we propose a multiuser MIMO detection method based on coding theoretical viewpoint. From (\ref{eq:obs}), the observation at the $i$-th receiver antenna after ADC quantizer is given by
\begin{align}
r_i & = \mbox{sign}\left(\hv^{\transp}_i \Mc(\mv) + z_i\right) \in \{0,1\}, \label{eq:o_channel}
\end{align} for $i \in [n_r]$, where $\mv=[m_1,\ldots,m_K]^{\transp}$. Then, our goal is to decode the users' messages $\hat{\mv}$ from the observations
 $\rv=[r_1,\ldots,r_{n_r}]^{\transp}$.
For this, we introduce an {\em equivalent communication model} from a coding theory perspective as illustrated in Fig.~\ref{e_model}. The proposed model consists of 
\begin{itemize}
\item {\em Encoding function} $\Ec$: This maps the users' messages $\mv$ into a codeword $\cv \in \Cc$ where notice that a code $\Cc$ is not designable but is completely determined as a function of the non-linear channel matrix;
\item {\em Effective channel}: This is composed of $n_r$ parallel BSCs with crossover probabilities $\epsilon_{i}(\mv)$ for $i \in [n_r]$. Note that the crossover probability depends on both the users' messages and the channel matrix;
\item {\em Decoding function} $\Dc$: This is what we will propose in this paper.
\end{itemize}

In this section, it is assumed that a channel matrix $\Hm$ is perfectly known at the BS. We start with the single user case of $K=1$ to present the main idea of the proposed coding method and then extend it into a general case in the sequel.
\subsection{Single User: Repetition Coding}\label{subsec:single}

Assuming that $K=1$, we specify an encoding function, an effective channel, and a decoding function of the equivalent communication model as illustrated in Fig.~\ref{e_model}. We define a code $\Cc=\{\cv_0,\cv_{1}\}$ as a function of a channel vector $\hv=[h_1,\ldots,h_{n_r}]^{\transp}$. Each codeword is defined as
\begin{equation}
\cv_m =\left[\mbox{sign}(h_1\Mc(m)),\ldots, \mbox{sign}(h_{n_r}\Mc(m))\right]^{\transp},\label{eq:1}
\end{equation} for $m \in \{0,1\}$.  This is nothing but a {\em repetition} code of length $n_r$ and rate $1/n_r$. Then, the input $\tv$ of an effective channel  is obtained by an encoding function $\Ec: \{0,1\}\rightarrow \Cc$ as
\begin{equation}
\tv=\Ec(m) = \cv_m \mbox{ for } m \in \{0,1\},
\end{equation}where $\cv_m = [c_{m,1},\ldots,c_{m,n_r}]^{\transp}$. In Fig.~\ref{e_model}, an effective channel consists of 
 $n_r$ parallel BSCs in which the crossover probability $\epsilon_i$ of the $i$-th sub-channel is computed from 
(\ref{eq:o_channel}) as
\begin{align}
\epsilon_i&= P(|h_i\sqrt{\mbox{SNR}}| + z_i < 0) = Q(|h_i\sqrt{\mbox{SNR}}|),
\end{align} for $i\in[n_r]$, where $Q(\cdot)$ denotes the Q-function as
\begin{equation}
Q(t) = \frac{1}{2\pi}\int_{t}^{\infty} \exp\left(-\frac{u^2}{2}\right) du.
\end{equation}
Using the above equivalent communication model, we transform a multiuser massive MIMO detection problem into the corresponding channel coding problem. Since a code $\Cc$ has been already generated, our goal is to design a decoding method. We first consider a {\em minimum distance} (MD) decoding because it has been widely used as the decoding method for repetition codes \cite{MacWilliams}. For MD decoding, a user's message is decoded as
\begin{equation}
\hat{m} = \argmin_{b\in\{0,1\}} d_h (\rv, \cv_b).
\end{equation} Next, we consider an optimal maximum likelihood (ML) decoding method. For ML decoding, a user's message is decoded as
\begin{equation}
\hat{m} = 
\begin{cases}
0 & \quad\text{if } P(\rv|m=0) > P(\rv|m=1)\\
1 & \quad\text{otherwise},
\end{cases}
\end{equation} where
\begin{equation}
P(\rv|m=b) = \prod_{i=1}^{n_r} ((1-\epsilon_i)\mathbf{1}_{\{r_i = c_{b,i}\}}+\epsilon_i \mathbf{1}_{\{r_i \neq c_{b,i}\}}),
\end{equation}where  $\mathbf{1}_{A}$ denotes an indicator function with $\mathbf{1}_{\{A\}}=1$ if $A$ is true, and $\mathbf{1}_{\{A\}}=0$, otherwise. 
%

To explain the difference between ML and MD decodings, we first provide the following definition.
\begin{definition}\label{def:w_MD}
For any two vectors $\av$ and $\bv$ of length $n_r$, we define a {\em weighted} Hamming distance $d_{wh}(\av,\bv)$  with the weights $\{\alpha_i\}_{i=1}^{n_r}$ and $\{\beta_{i}\}_{i=1}^{n_r}$ as
\begin{equation}
d_{wh}(\av,\bv) = \sum_{i=1}^{n_r} \alpha_i \mathbf{1}_{\{a_i = b_i\}} + \sum_{i=1}^{n_r} \beta_i \mathbf{1}_{\{a_i \neq b_i\}}.
\end{equation}
\end{definition}

\begin{figure}
\centerline{\includegraphics[width=9cm]{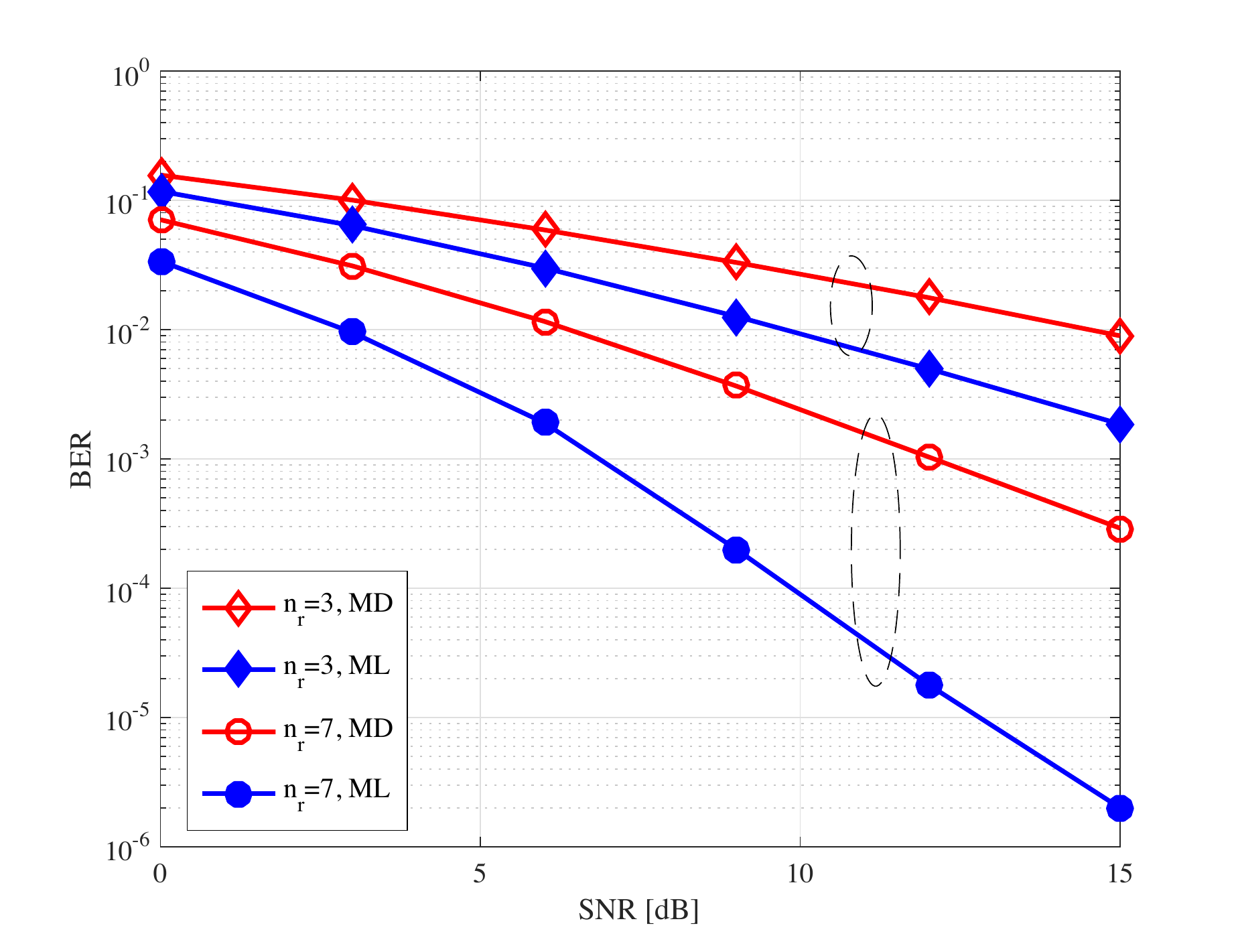}}\vspace{-0.5cm}
\caption{$K$=1. Performance comparison of ML and MD decoding methods.}
\label{SIMO-DEC}
\end{figure}

The following Remark~\ref{remark:MD_ML} explains the reason why ML decoding should be used for our problem, while MD decoding is optimal in a classical coding problem.  
\begin{remark}\label{remark:MD_ML} {\em (MD vs. ML)} From Definition~\ref{def:w_MD},  both ML and MD decoding methods  can be represented with an unified view as
\begin{equation}
\hat{m} = \argmin_{b\in\{0,1\}} d_{wh} (\rv, \cv_b),
\end{equation} with the different weights. Specifically, the weights when using ML decoding are $\alpha_i = -\log{(1-\epsilon_i)}$ and $\beta_i = -\log{\epsilon_i}$ for $i \in [n_r]$. The weights when employing MD decoding are $\alpha_i = 0$ and $\beta_i = 1$ for $i \in [n_r]$. 

It would be reasonable to allocate a higher belief (or weights) for the information provided by more reliable channels. On the one hand, ML decoding assigns proper soft-weights according to the channel reliabilities. On the other hand, MD decoding assigns hard weights, which does not contain the channel reliabilities adequately. Because of this difference, ML decoding outperforms MD decoding as shown in Fig.~\ref{SIMO-DEC}. Then, a natural question aries: {\em why MD decoding is used in a classical coding problem?} This is because MD decoding is equivalent to ML decoding when all sub-channels have the same reliabilities (i.e., $\epsilon_i = \epsilon$ for $i \in [n_r]$). That is, in a classical coding problem, a codeword of length $n$ is transmitted over a statistically equivalent channel (i.e., $\epsilon$ is unchanged) and hence, MD decoding is optimal. Whereas, this is not the case in our problem and MD decoding is highly suboptimal. From Fig.~\ref{SIMO-DEC}, we observe that ML decoding can achieve a higher diversity gain over MD decoding, i.e., the performance gap is unbounded as $\mbox{SNR}$ increases. Nevertheless, ML decoding has a possible advantage in terms of implementation complexity  because it is only required to know a code $\Cc$,  while ML decoding additionally needs to know the channel reliabilities of all sub-channels. To resolve this issue, we present a practical ML decoding method in Section~\ref{sec:practical}.
\hfill$\blacksquare$
\end{remark}

\subsection{$K$ Users: Non-linear Coding}

In this case, each user $k$ transmits its message $m_k \in \{0,1\}$ to the BS. As explained before, we define an {\em effective} channel that consists of $n_r$ parallel BSCs, whose input and output are  $\tv=[t_1,\ldots,t_{n_r}]^{\transp}$ and $\rv=[r_1,\ldots,r_{n_r}]^{\transp}$, respectively.  The $i$-th sub-channel has the crossover probability $\epsilon_{i}(\mv)$ as
\begin{equation}
\epsilon_i(\mv) = Q(|\hv_i^{\transp}\Mc(\mv)|),\label{eq:cross}
\end{equation}  for $i \in [n_r]$. Notice that the $\epsilon_{i}(\mv)$ depends on the users' messages $\mv$ as well as a channel matrix. 

Define a {\em channel-dependent} code $\Cc=\{\cv_0,\ldots,\cv_{2^{K}-1}\}$ with
\begin{equation*}
\cv_k= \left[\mbox{sign}\left(\hv_{1}^{\transp}\Mc(g(k))\right),\ldots, \mbox{sign}\left(\hv_{n_r}^{\transp}\Mc(g(k))\right)\right]^{\transp}.
\end{equation*} Notice that the minimum distance of $\Cc$ is completely determined as a function of $\Hm$ and ADC quantization. Thus, we let $d_{{\rm min}}(\Hm)$ denote the minimum distance of $\Cc$ generated by the channel matrix $\Hm$ and the ADC quantization. The input $\tv$ of an effective channel is obtained by an encoding function $\Ec$ as
\begin{equation}
\tv=\Ec(\mv) = \cv_{g^{-1}(\mv)}.
\end{equation}
As in the single-user case, we consider two decoding methods: 
\begin{enumerate}
\item MD decoding finds users' messages $\hat{\mv}=g(\hat{k})$ from
\begin{equation}
\hat{k} = \argmin_{k \in \{0,1,\ldots,2^{K}-1\}} d_{h}(\rv, \cv_{k}).\label{eq:MDD}
\end{equation} 
\item ML decoding finds users' messages $\hat{\mv}=g(\hat{k})$ from
\begin{equation}\label{eq:wMD}
\hat{k} = \argmin_{k \in \{0,1,\ldots,2^{K}-1\}} d_{wh}(\rv, \cv_{k}),
\end{equation} with the weights $\alpha_i = -\log{\left(1-\epsilon_i(g(k)\right)} $ and $\beta_i = -\log{\epsilon_i(g(k))}$ for $i \in [n_r]$, where $\epsilon_i(g(k))$ is in (\ref{eq:cross}).
\end{enumerate}


\begin{figure}
\centerline{\includegraphics[width=9cm]{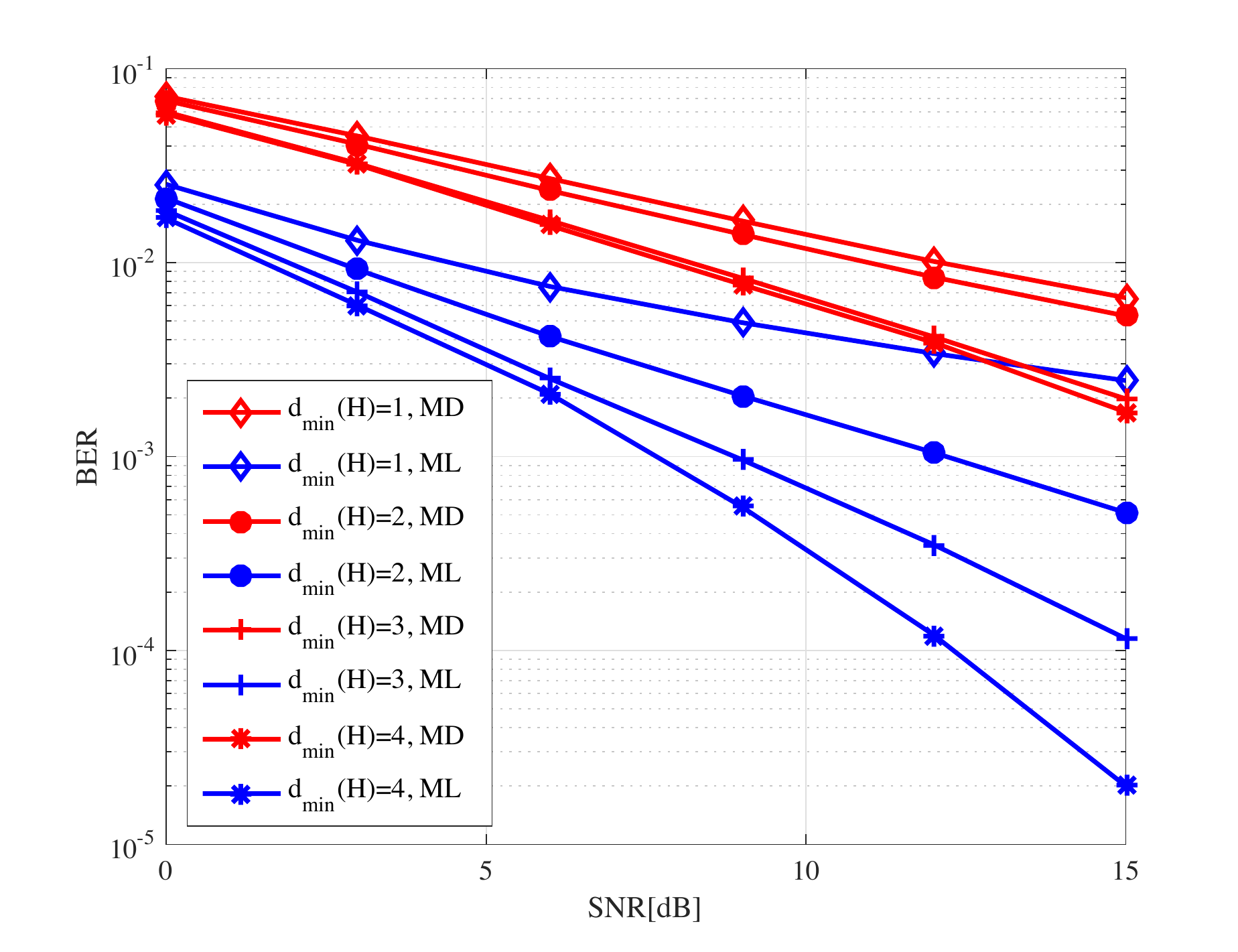}} \vspace{-0.5cm}
\caption{$K=4$ and $n_r=16$. The receiver diversity gain as a function of a minimum distance of $\Cc$ (i.e., $d_{{\rm min}}(\Hm)$).}
\label{MIMO_MD} 
\end{figure}

\begin{figure}
\centerline{\includegraphics[width=9cm]{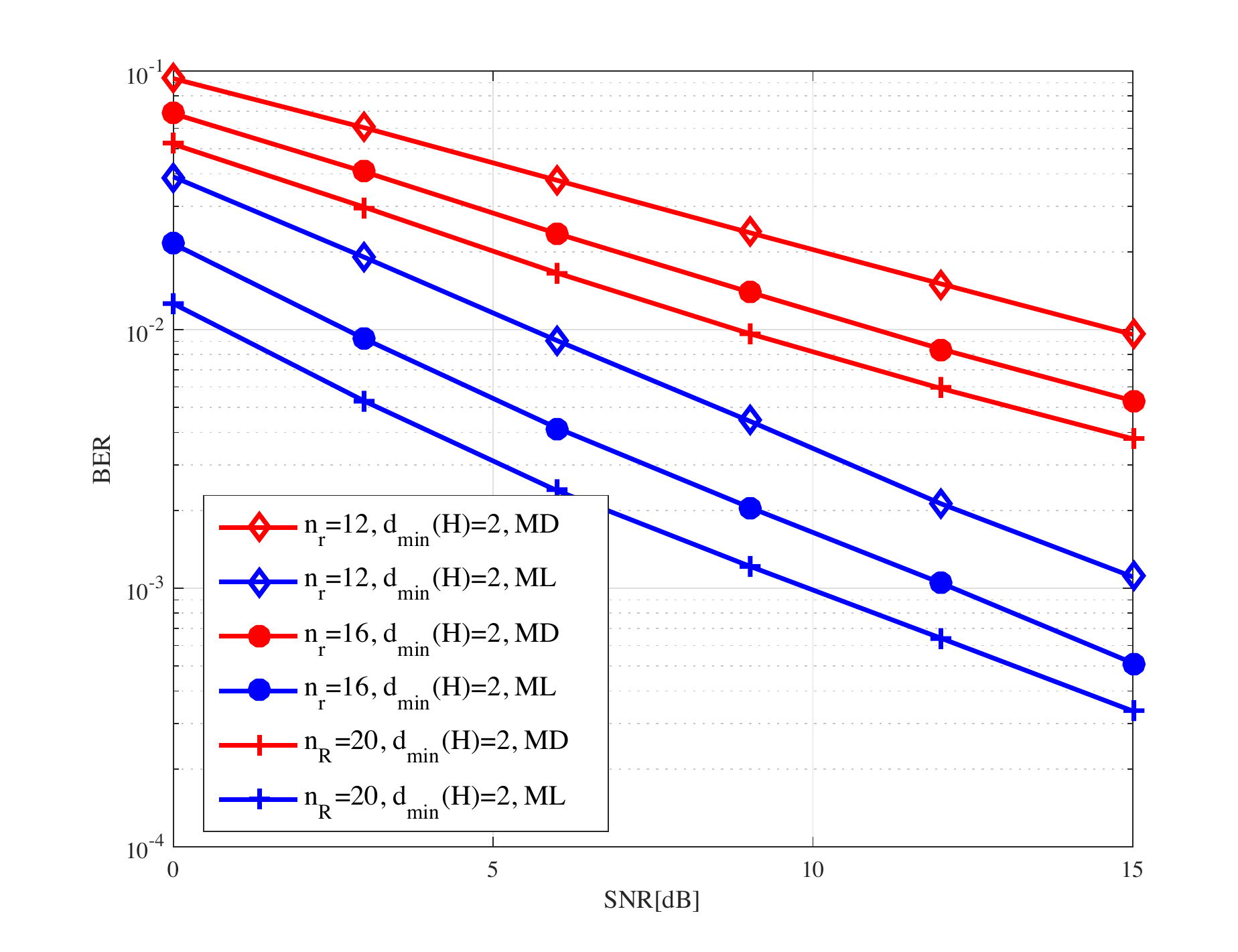}}\vspace{-0.5cm}
\caption{$K=4$. The performance comparison of the proposed coding method as a function of $n_r$ when a random channel matrix yields the same minimum distance of the associated code $\Cc$.}
\label{power_gain}
\end{figure}

\begin{remark}\label{remark:diversity} {\em (Diversity Order)} Suppose the uplink multiuser massive MIMO system with $K=4$ and $n_r = 16$ in which a random channel matrix $\Hm \in \RR^{16 \times 4}$ is used and its element follows an IID Gaussian random variables with zero-mean and unit-variance. Let $\Hc$ denote the sample space containing all possible channel realizations. Let $\Ac_{\ell} = \{\Hm \in \Hc: d_{{\rm min}}(\Hm) = \ell\} \subseteq \Hc$ denote the subset of all channel realizations such that the corresponding codes have the minimum distance $\ell$. To see the impact of minimum distance on the performance, we consider the conditional error probabilities defined as $P_e(d_{{\rm min}}(\Hm)=\ell)=\frac{1}{K}\sum_{k=1}^{K} P(\hat{m}_k \neq m_k | \Ac_\ell)$. The corresponding numerical results are provided in  Fig.~\ref{MIMO_MD}. From this, we can see that for ML decoding, the diversity order is strongly related to the minimum distance of $\Cc$ (i.e., $d_{{\rm min}}(\Hm)$), since the slope of BER curve is enhanced as $d_{{\rm min}}(\Hm)$ increases. For MD decoding, however, the diversity order seems to be related to the error-correction capability defined as $\lfloor \frac{d_{{\rm min}}(\Hm) - 1}{2} \rfloor$ \cite{MacWilliams}. Fig.~\ref{power_gain} shows that increasing the number of receiver antennas without improving a minimum distance only attains a SNR gain, i.e., the slope of BER curve is unchanged. Thus, it would be interesting to investigate an user selection algorithm such that the minimum distance of a resulting code $\Cc$ is maximized, which is left for a future work.
\hfill$\blacksquare$
\end{remark}


\section{Practical Implementation of \\the Proposed Method}\label{sec:practical}

In Section~\ref{sec:Main}, we have presented a multiuser MIMO detection method using coding theory, by assuming that BS completely knows a channel matrix $\Hm$. In practice, however, it is difficult to estimate the channel matrix perfectly due to non-linear distortion effects by the ADCs \cite{Choi}. Instead of estimating of a channel matrix $\Hm$, we directly estimate a code $\Cc$ (i.e., $2^K$ codewords of $\Cc$) using training sequences in this section.

We consider a block fading channel in which the channel is static for $N$ channel uses in a given fading block and changes independently from block-to-block. Let the first $2^{K}T<N$ channel uses be devoted for a training phase and the remaining $N-2^K T$ channel uses be dedicated to a data communication phase. 
During the overall $2^K T$ time slots, we estimate the codewords $\cv_{0},\cv_{1},\ldots,\cv_{2^K -1}$ in that order. To estimate a codeword $\cv_k \in \Cc$, each user $i$ transmits a training sequence $[b_{i-1},b_{i-1},...,b_{i-1}]$ of length $T$ over the $T$ time slots, where  $k=b_02^0+b_12^1+\cdots+b_{K-1}2^{K-1}$. 
Then, the corresponding observations at the $i$-th receiver antenna at a BS is denoted by 
\[\rv_{i}^{(k)}=[r_{i,kT+1},\ldots,r_{i,(k+1)T}]^{\transp}\] for $i \in [n_r]$. From this, we estimate the $i$-th element of $\cv_k$, $c_{k,i}$ using the simple majority decoding as
\begin{equation}
\hat{c}_{k,i} = 
\begin{cases}
0 & \mbox{ if } d_w(\rv_{i}^{(k)}) < T/2\\
1 & \mbox{ otherwise}
\end{cases}
\end{equation} for $i \in [n_r]$. Repeating the above procedures for $k=1,\ldots,K$, we can estimate a code $\Cc=\{\hat{\cv}_0,\ldots,\hat{\cv}_{2^K-1}\}$ where $\hat{\cv}_k = [\hat{c}_{k,1},\ldots,\hat{c}_{k,n_r}]^{\transp}$. Then, MD decoding can be performed using the estimated code $\Cc$.

In Section~\ref{sec:Main}, an important observation is that ML decoding can attain a non-trivial gain over MD decoding by leveraging different channel reliabilities of sub-channels. In practice, however, it is quite complicated to exactly estimate $\epsilon_{i}(\mv)$. As a practical implementation of ML decoding, we propose a {\em weighted} MD decoding using the weighted Hamming distance in Definition~\ref{def:w_MD}. Here, users' messages $\hat{\mv} = g(\hat{k})$ are decoded as
\begin{equation}\label{eq:w_MD}
\hat{k}= \argmin_{k \in \{0,1,\ldots,2^{K}-1\}} d_{wh}(\rv, \hat{\cv}_{k}),
\end{equation} with the estimated weights 
\begin{align}
\hat{\alpha}_i &= -\log{\left(1-\frac{1}{T}\sum_{t=1}^{T}d_h(\hat{c}_{k,i},r_{i,(kT+t)})\right)}\\
\hat{\beta}_i &= -\log\left({\frac{1}{T}\sum_{t=1}^{T}d_h(\hat{c}_{k,i},r_{i,(kT+t)})}\right)
\end{align} for $i \in [n_r]$.  This decoding method can be understood that an empirical error probability of each sub-channel is used to capture a channel reliability. Also, as $T$ grows, the performance of weighted MD decoding is close to that of ML decoding. For the case of a small training overhead, an empirical error probability can be zero although the corresponding sub-channel should not be a perfect channel. To overcome this problem, a minimum value of an empirical error probability is used. Note that the performance is not sensitive to a minimum value, if it is small enough (e.g., $10^{-7}$).

\begin{figure}
\centerline{\includegraphics[width=9cm]{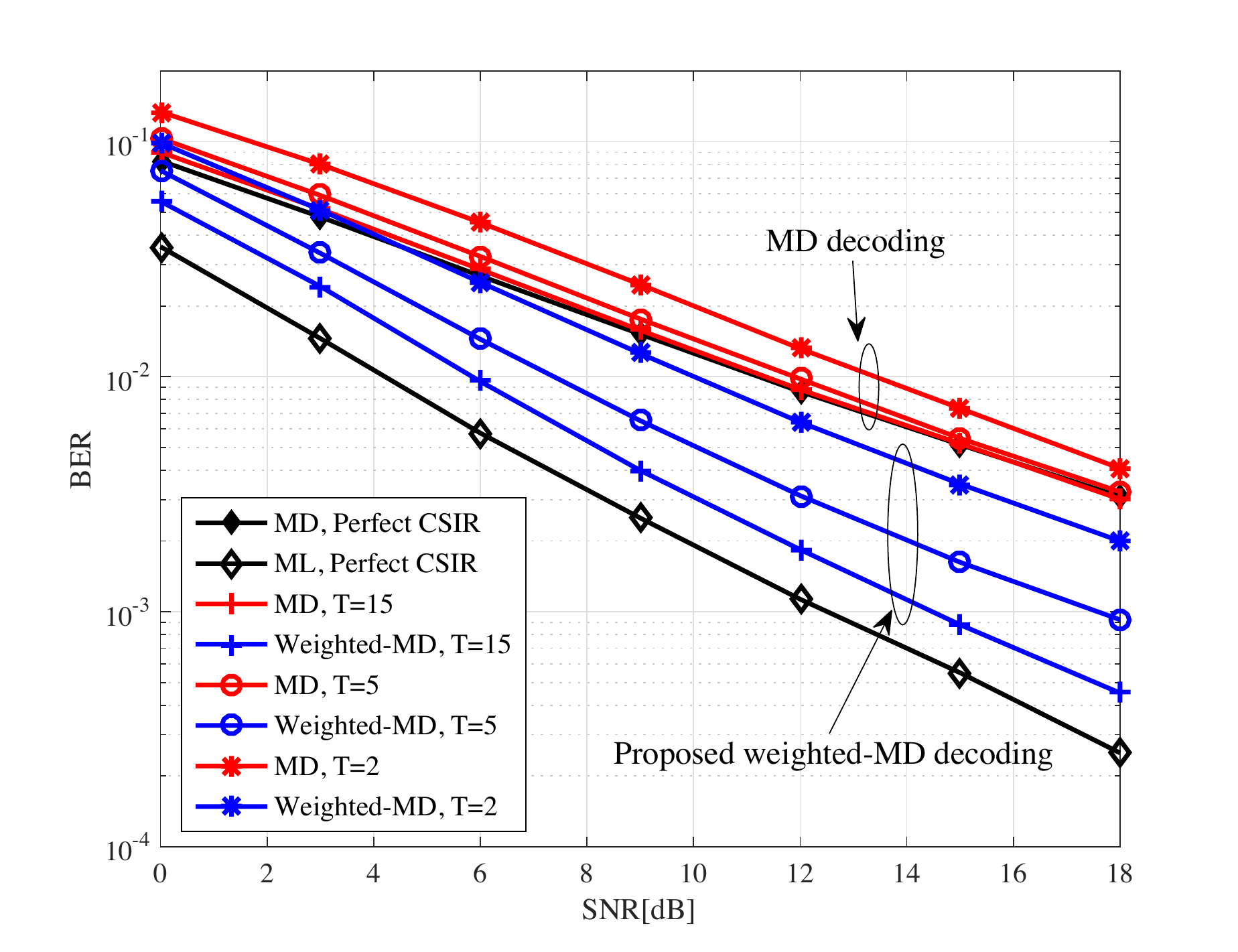}}\vspace{-0.5cm}
\caption{$K=2$ and $n_r = 9$. The BER performance of the proposed coding method as a function of training overhead.}
\label{training}
\end{figure}

As an example, suppose the multiuser massive MIMO system with $K=2$ and $n_r=9$ where a random channel matrix is generated following the same method  in Remark~\ref{remark:diversity}. Fig.~\ref{training} shows that weighted MD decoding attains a non-trivial gain over MD decoding, even for a lower training overhead. As expected, the performance gap increases as the training overhead grows.

\section{Conclusion}

We proposed a novel multiuser MIMO detection method. The proposed method is to convert the multiuser detection problem to an equivalent coding problem by leveraging coding theory. One major implication of the proposed method is that  the minimum distance of a channel-dependent code is strongly related to a diversity order. We further presented a ML decoding method that does not require CSIR at the BS using a training sequence.

\section*{Acknowledgement}

This work was supported in part by the Electronics and Telecommunications Research Institute through the Korean Government (Wireless Transmission Technology in Multi-point to Multi-point Communications) under Grant 16ZI1100.



\end{document}